\documentclass[%
 reprint,
 amsmath,amssymb,
 aps
floatfix,
]{revtex4-1}

\usepackage{color}
\usepackage{graphicx}
\usepackage{dcolumn}
\usepackage{bm}
\usepackage{subfigure}
\usepackage{caption}
\usepackage{natbib}
\usepackage[bookmarks=false]{hyperref}
\begin{document}

\preprint{APS/123-QED}

\title{Nb Implanted BaO as a Support for Gold Single Atoms
}

\author{Debolina Misra}
 \email{mm18ipf03@iitm.ac.in}
 \author{Satyesh K. Yadav}
\affiliation{
Department of Metallurgical and Materials Engineering, Indian Institute of Technology Madras, Chennai, 600036, India}

\date{\today}

\begin{abstract}
Using first-principles modelling based on density functional theory we show that oxides implanted with transition metal can act as support for Au single atoms, which are stable against agglomeration. In our previous work we have shown that implanted transition metal, doped in BaO is stable as interstitial in various charge states by transferring the excess charge to an acceptor level close to VBM. Taking Nb as an example we show that single atom Au has its Fermi level close to the VBM of BaO and hence is able to accept charge from the dopant. This charge transfer process between Nb and Au helps Au atoms bind strongly on the doped BaO support. We also show that these charged Au atoms repel each other and prefer to remain atomically dispersed preventing cluster formation. Substitutional doping of transition metals have earlier been reported to bind Au atoms. However, if doped at interstitial sites, they can bind more Au atoms; for example, 5 Au atoms can be anchored per Nb dopant present in BaO interstitial, compared to 3 Au atoms when Nb is doped at substitutional site. This work paves the way for an altogether new technique of stabilizing noble metal single atoms on transition metal doped oxides.
\end{abstract}

\maketitle


\section{\label{sec:intro}Introduction} 
In spite of the tremendous growth in designing non-noble materials for catalysis, noble metals are still at the heart of catalysis research due to their highest catalytic activity \cite{Liu2017}. Noble metal nanoparticles (NPs) are often used due to the large surface-to-volume ratio and the higher concentration of under coordinated-surface sites they expose, compared to larger particles with more extended surfaces. 
Nevertheless, the high cost and low abundance of noble metals still significantly hinder the use of such NPs as catalysts \citep{AB_AChem, zhang2014}. Single-atom catalysts (SACs), which consist of atomically dispersed metal atoms on a support, are promising solutions to reduce this cost. They maximize the surface-to-volume ratio, which can potentially lead to a way more efficient use of the noble metals and thus decrease the cost of catalyst fabrication \cite{Cheng2016,yang2013}. Despite existing challenges in preparation of stable and active SACs, the last decade has seen several successful strategies for fabricating promising SAC prototype \cite{sun_scirep, Qiao2011, Jones150}.

Nevertheless, atomically dispersed metals are often unstable against agglomeration into larger metal nanoparticles \cite{liu_jacs,chen2014,Jones150}. Efficient SACs must therefore be catalytically active and resistant to sintering, both of which strongly depend on intricate metal-support interactions. Doped oxides serve as very effective supports for catalysts. The dopant first changes the electronic structure of the host and can cause a charge transfer which is not only limited to the direct interaction of the dopant and its nearby host atoms, but can also be extended to the adsorbate, situated on the surface of the host lattice \cite{shao_angchem}. Such processes of charge transfer between impurity, the host lattice and the adsorbate are at the heart of heterogeneous catalysis where charge transfer to a proper adsorbate on the oxide surface can reduce the barrier of a chemical reaction. Recently it has been shown that for a charge transfer to take place between a dopant and an adsorbate, a direct interaction is not always necessary \cite{prada2013}. Some earlier studies have revealed that adsorption properties of gold and its growth pattern strongly depends on charge transfer to gold atoms which in turn depends on the nature of the dopant and the host lattice \cite{Sterrerprl,ricciprl}. For example, Mo transfers charge to Au in CaO and MgO and alters its growth pattern on the oxide surface. However, presence of Cr in both the oxides resulted in no charge transfer between gold atom and Cr \cite{Stavale2012}. 

To stabilize single atom Au, all the previous studies have focused on substitutional doping of oxides with transition metals (TMs). Here we conclusively show that interstitially doped oxides can act as better support as TMs at interstitial site can transfer more charge than at substitutional site. We use density functional theory (DFT) to show: (1) Nb prefers to be in 5+ charge state and occupy the interstitial site in BaO, (2) doped Nb transfers excess electron to Au atoms adsorbed on the surface and binds them strongly, and 3) Au atoms anchored to the support repel each other and resist cluster formation.

\section{\label{sec:method}Methodology}
All the spin-polarized DFT calculations were performed employing projector-augmented wave (PAW) method \cite{blochl1994projector} and a plane wave basis set with 500 eV energy cut-off, as implemented in Vienna \textit{Ab initio} Simulation Package (VASP) \cite{kresse1996efficient, kresse1996commat}. Generalized gradient approximation (GGA) was used to treat electronic exchange and correlation, employing the Perdew, Burke, and Ernzerhof (PBE) functional \cite{perdew1996generalized}. A \textit{k}-point mesh of 4x4x4 was used for achieving converged results within 10$^{-4}$ eV per atom. All the structures were fully relaxed using the conjugate gradient scheme and relaxations were considered converged when forces on each atom was smaller than 0.02 eV/\AA. Calculation of density of states (DOS) were performed using linear tetrahedron method with Blochl corrections \cite{blochlprb} and a denser k-grid. For the bulk calculations, cubic super cell containing 32 formula units of BaO was doped with Nb in all their possible charge states which results in a dilute limit(3.1\%) of doping.
\section{Results}
\subsection{Stability and preferred charge state of Nb dopant in BaO}
First we attempt to understand charge state of Nb when doped in BaO and its electronic structure, to ascertain its potential to transfer charge to Au. Ionic radii of Nb in 6 coordinated environment are 0.70, 0.68 and 0.64 Å, in +3, +4 and +5 charge states, respectively; while ionic radius of Ba is 1.35 Å.  Our prediction model \cite{sci_rep} suggests that Nb will be stable at interstitial in bulk BaO. We indeed find that Nb is stable at interstitial site. We calculate defect formation energy $E_f^q$ \cite{Rampiprl, Freysoldtprb2016, VandeWalleRev2014, BAJAJ2015} of Nb in bulk BaO as a function of electronic chemical potential $\mu$, using the following equation.
\begin{equation}
E_f^q=E_D^q-E_B-\eta+q(\mu+E_{ref}+{\Delta}V)+E_{corr}^q
\end{equation}
Here $E_D^q$ and $E_B$ are the total energies of the defect supercell with charge q and the defect free host supercell, respectively. $\eta$ is the chemical potential of the transition metal atom species. The '-' sign indicates the addition of defect in the host. $E_{ref}$ is a suitable reference energy, taken to be the valence band maximum (VBM), and $\Delta$V is the correction to realign the reference potential of the defect supercell with that of the defect free supercell \cite{VandeWallejap2004}. $E_{corr}^q$ is the correction to the electrostatic interaction and the finite size of the supercell. In this work only the first-order monopole correction has been taken into account.  
\begin{figure}[t!]
\centering
\includegraphics[height=1.8in]{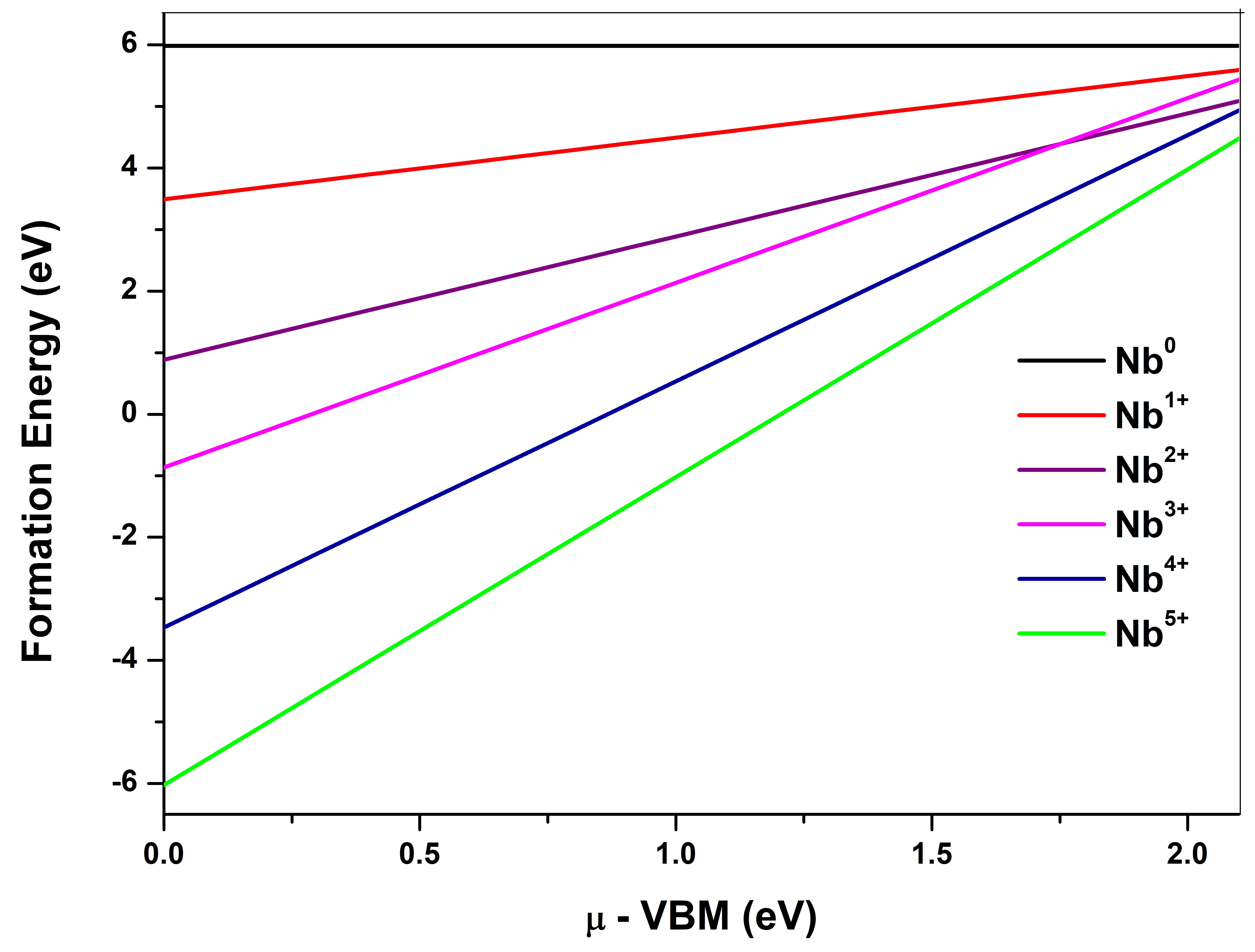}
\caption{Formation energy for the neutral and charged Nb interstitial in bulk BaO as a function of electronic chemical potential $\mu$.} 
{\label{bao_form}}
\end{figure}

Defect formation energy of Nb in bulk BaO as function of electronic chemical potential is shown in Fig.\ref{bao_form}. Nb is stable in a single charge state(5+) for the entire range of electronic chemical potential ($\mu$) studied. Here $\mu$ varies from VBM up to the band-gap of the host oxide, obtained from our DFT calculation. As we intend to use Nb doped BaO as a support for single atom Au, we explored the followings : 1) change in Fermi-level position of doped BaO with different charge states of Nb and 2) stability of Nb in BaO at various depth from the surface.

\begin{figure}[t!]
    \centering
    \begin{subfigure}
        \centering
        \includegraphics[height=1.2in]{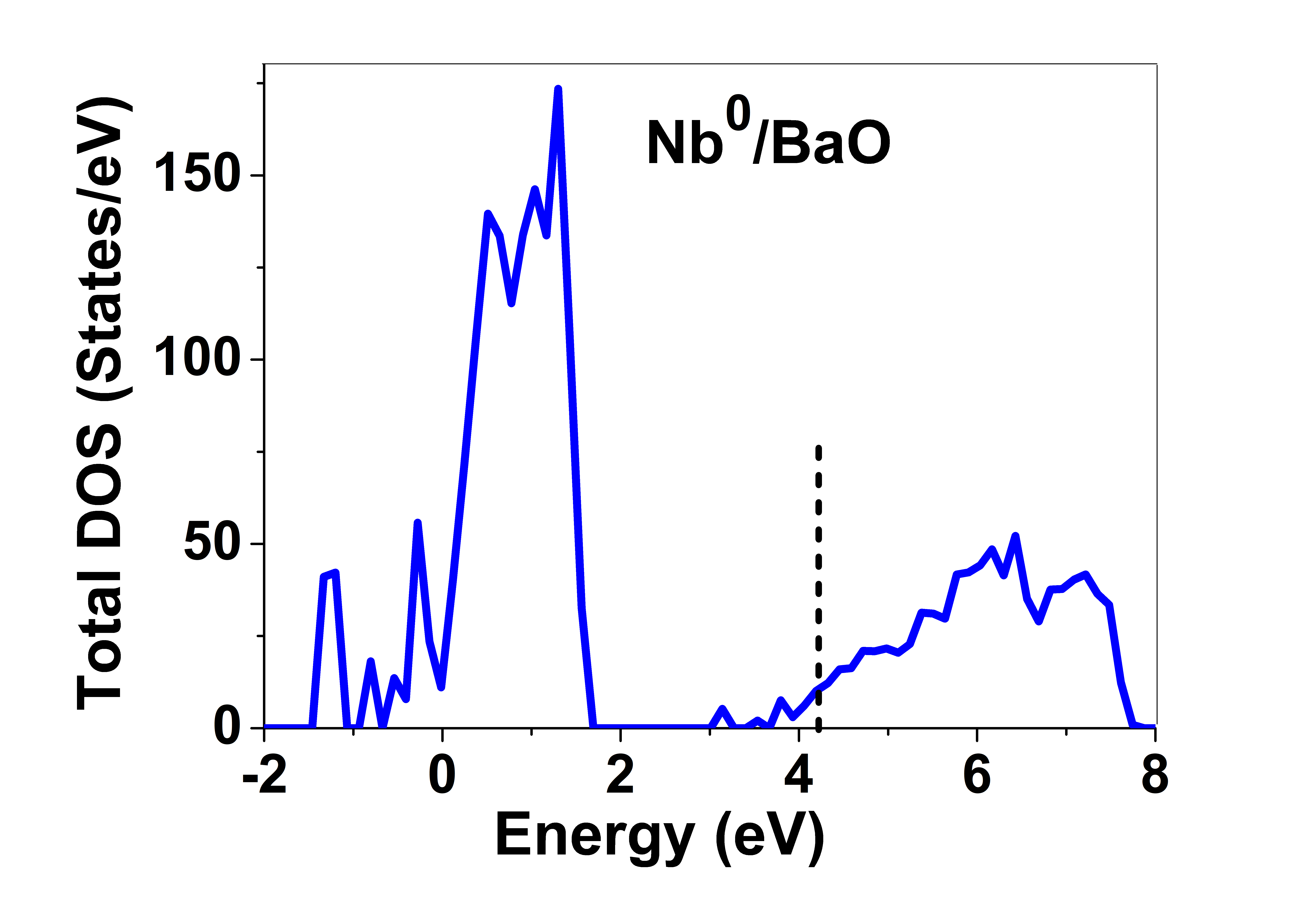}
    \end{subfigure}
    \vspace{-0.2em}
    \begin{subfigure}
        \centering
        \includegraphics[height=1.2in]{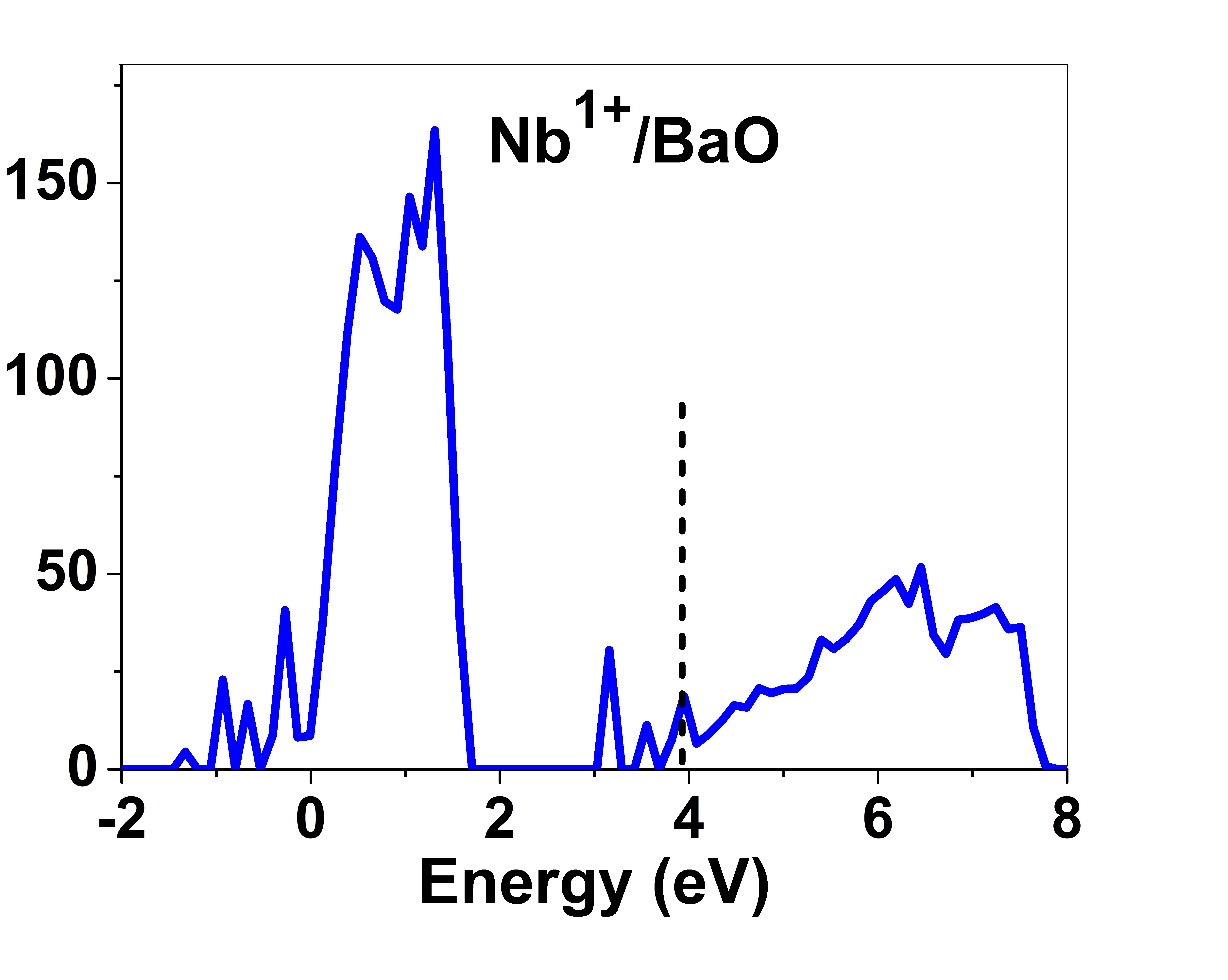}
    \end{subfigure}
    \vspace{-0.2em}
    \begin{subfigure}
        \centering
        \includegraphics[height=1.2in]{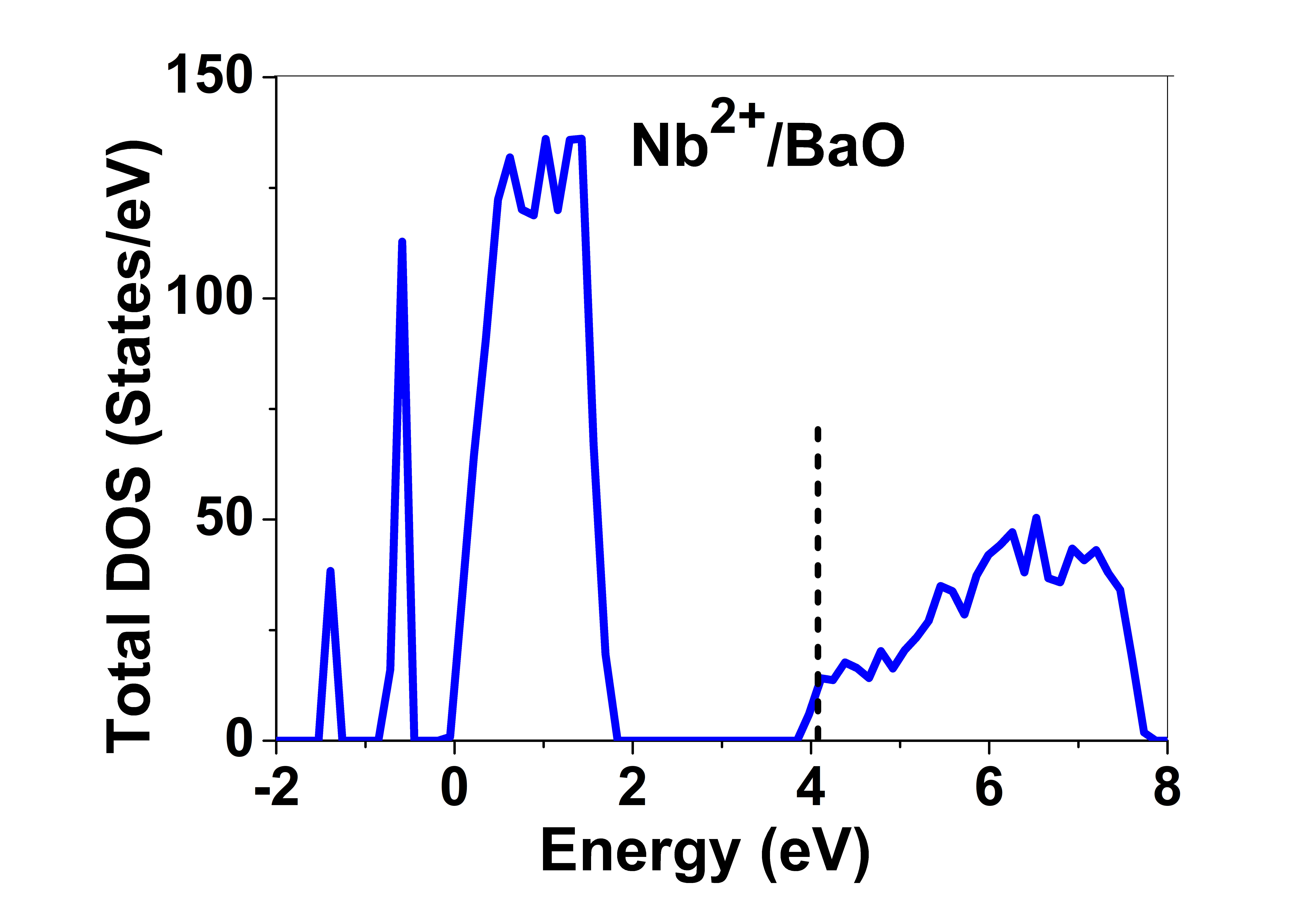}
    \end{subfigure}
    \vspace{-0.2em}
    \begin{subfigure}
        \centering
        \includegraphics[height=1.2in]{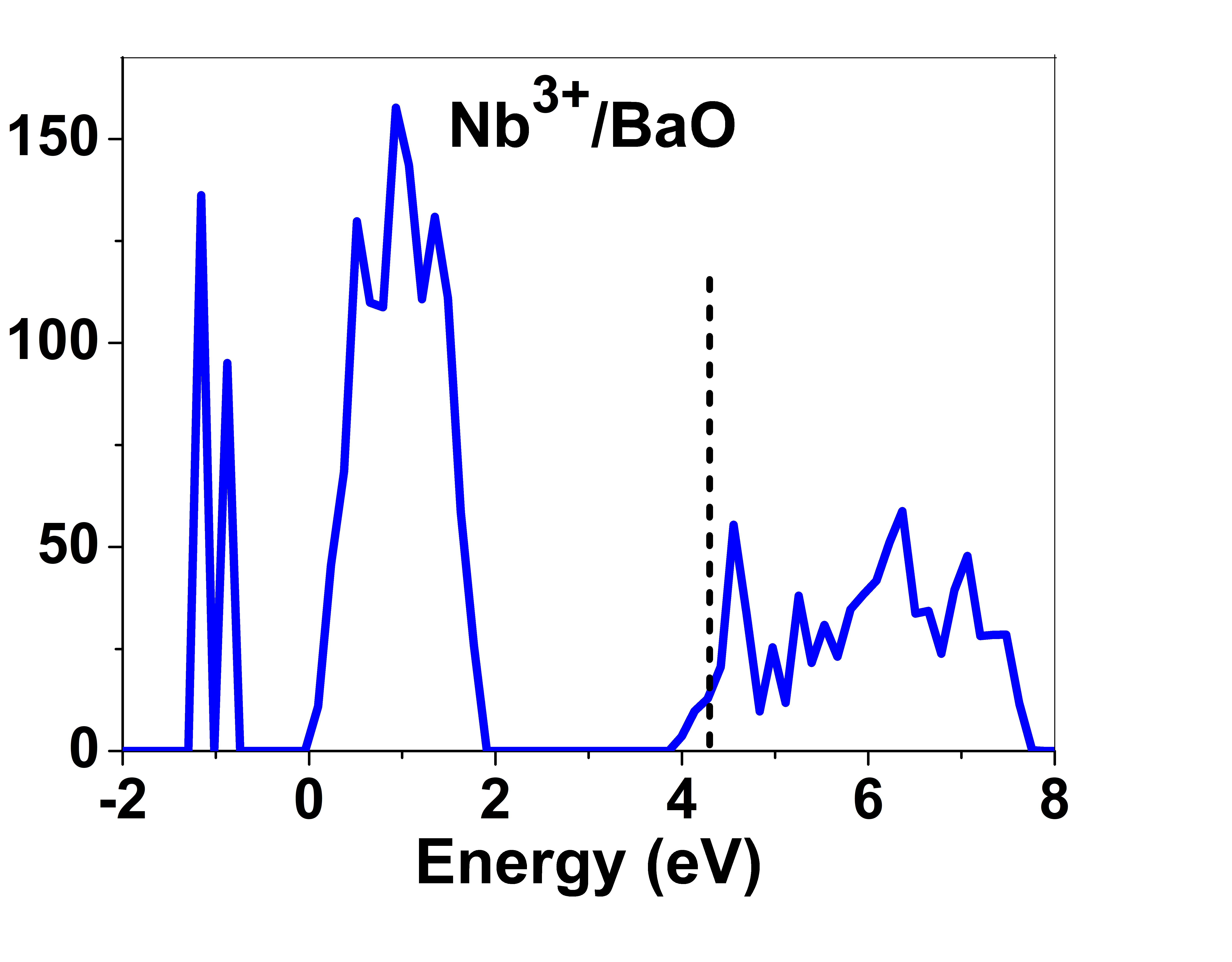}
    \end{subfigure}
    \vspace{-0.2em}
    \begin{subfigure}
        \centering
        \includegraphics[height=1.2in]{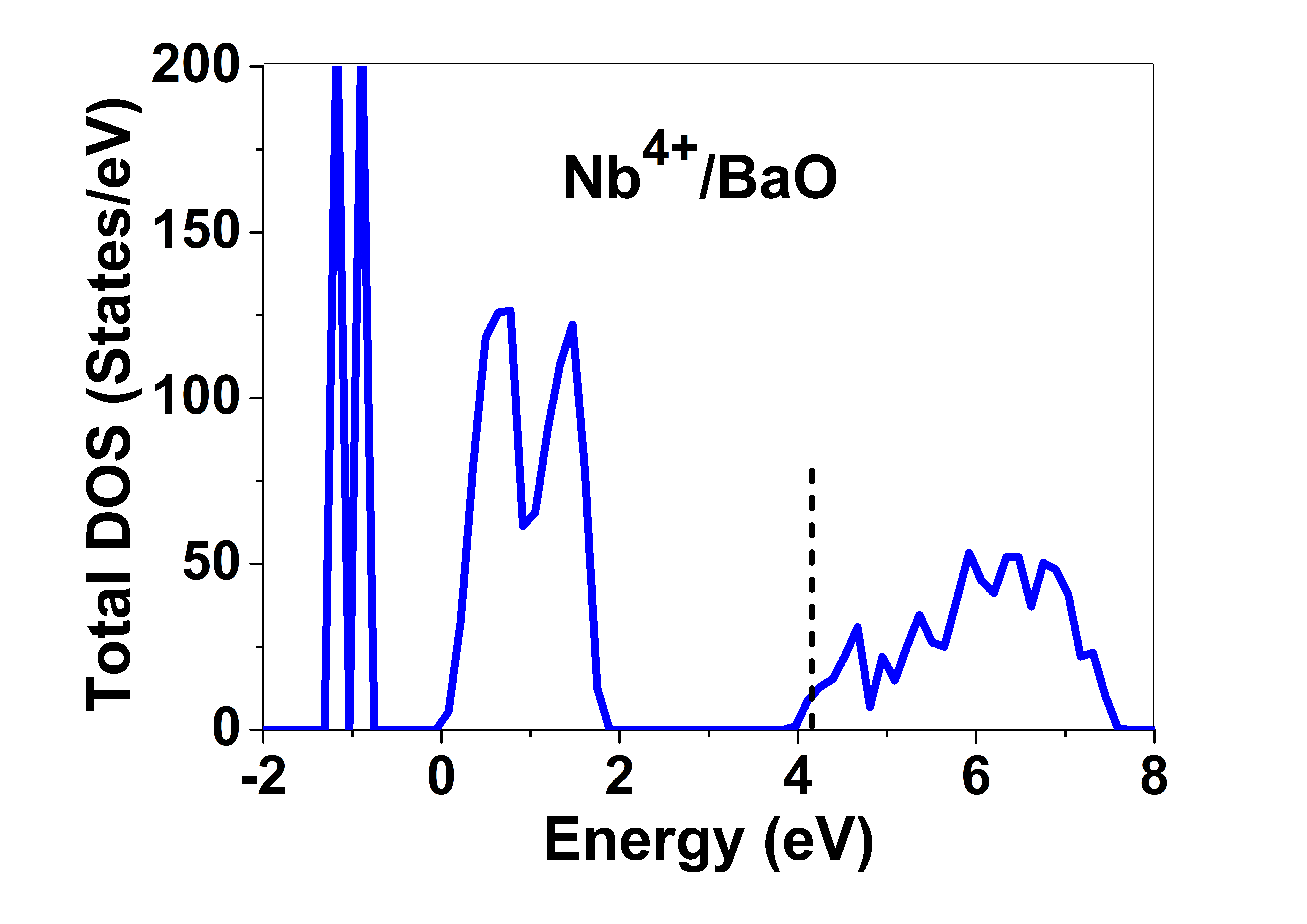}
    \end{subfigure}
    \vspace{-0.2em}
    \begin{subfigure}
        \centering
        \includegraphics[height=1.2in]{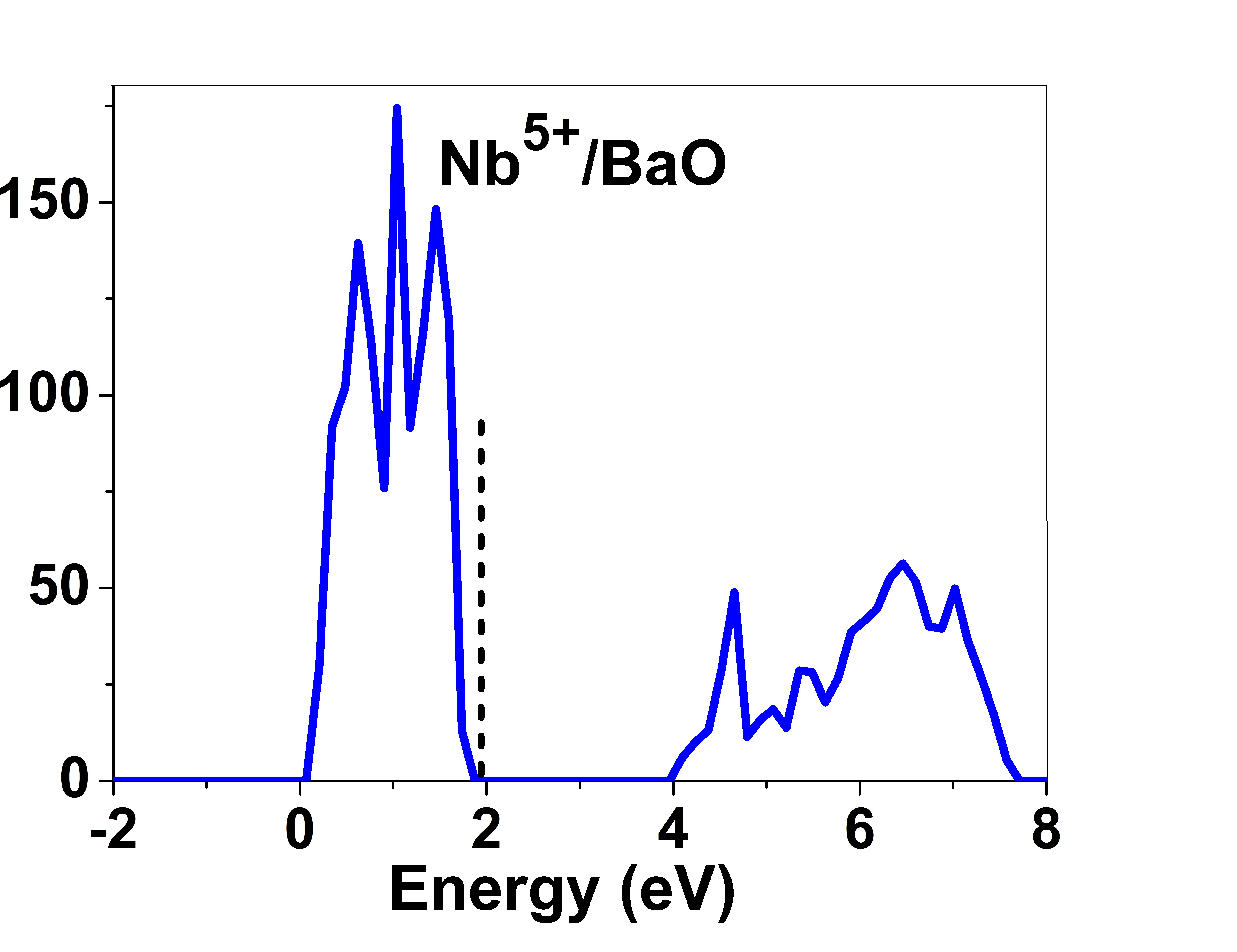}
    \end{subfigure}
    \vspace{-0.2em}
    \begin{subfigure}
        \centering
        \includegraphics[height=1.1in]{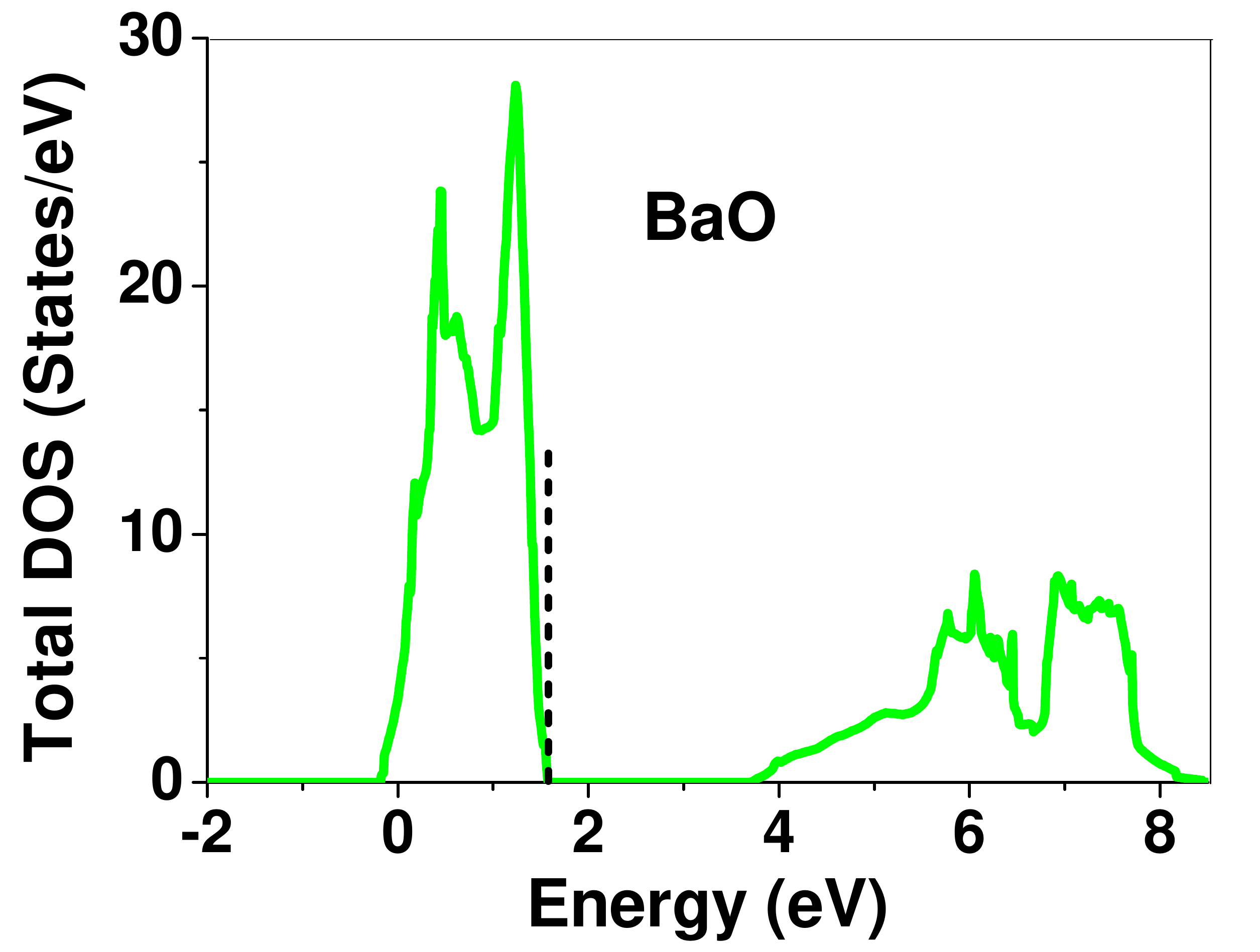}
    \end{subfigure}
    \caption{Total density of states of Nb-doped BaO and pure BaO. The vertical dashed line indicates the Fermi level.} 
{\label{bao_dos}}
\end{figure}

Density of states for Nb-doped bulk BaO is shown in Fig.\ref{bao_dos}. For all the charge states of Nb except 5+, the Fermi level lies in the antibonding state which indicates that these states are not stable. The Fermi level lies at the VBM only for 5+ charge state of Nb in BaO which suggests that this state is the most stable one. Change in Fermi-level position for different charge states of Nb is similar to substitutionally doped oxides, generally used to stabilize single atoms \cite{prada2013, Stavale2012}. We can expect that the five valence electrons of Nb can be transferred to an acceptor when BaO is doped with Nb.

Stability of Nb interstitials at various depth from the surface of BaO has been investigated in order to trace its most preferred position. A 5 layer thick BaO slab containing \{001\} surfaces (known to be one of the most stable surfaces of BaO) is used to model the surface. In-plane dimension of supercell is 2x2 units of BaO with a 12 Å vacuum along [001] direction. Atoms in the bottom two layers of the slab is frozen to their bulk positions in all our calculations.
\begin{figure}[t!]
\centering
\includegraphics[height=2.5in]{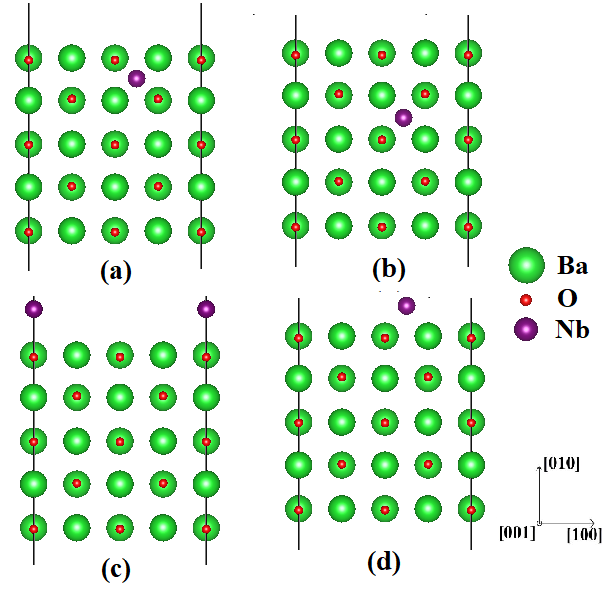}
\caption{BaO\{001\} surface with Nb in different layers; Nb at (a) subsurface (b) sub-subsurface and on surface (c) O-top and (d) hollow sites respectively.} 
{\label{nb_diffz}}
\end{figure}

Nb in different layers (subsurface, sub-subsurface and on surface O-top and hollow sites) of BaO\{001\} surface is shown in Fig.\ref{nb_diffz}. The relative energies between all these three configurations for both neutral Nb and Nb in 5+ charge state are listed in Table\ref{tab:nb_z}.

\begin{table}
\setlength{\tabcolsep}{4pt}
  \begin{center}
    \caption{Relative energy of BaO \{001\} surfaces containing Nb at different layers}
    \label{tab:nb_z}
     \begin{tabular}{c|c|c}
      \multicolumn{1}{c|}{Position} & \multicolumn{2}{c}{Relative energy (eV)}\\
        & \multicolumn{1}{c|}{Nb$^0$} & \multicolumn{1}{c}{Nb$^{5+}$}\\
      \hline
      Subsurface (Fig.\ref{nb_diffz}a) & 0.00 & 0.00\\
      Sub-subsurface (Fig.\ref{nb_diffz}b) & 1.76 & 1.22 \\
      On surface O-top (Fig.\ref{nb_diffz}c)& 0.80 & 9.58\\
      On surface Hollow (Fig.\ref{nb_diffz}d) & 1.50 & 6.09 \\
   \end{tabular}
  \end{center}
\end{table}

From Table\ref{tab:nb_z} it can be seen that subsurface layer is energetically the most preferred layer for Nb interstitial in both neutral and 5+ charge states. However, the most important observation one can make from this calculation is, although Nb neutral is stable in the subsurface layer, the difference between configurations with Nb occupying the subsurface interstitial and on surface O-top site is very small (0.18 eV). 
Hence under high temperature, neutral Nb can easily escape from the subsurface layer and can occupy the surface O-top site. Nb occupying surface O-top site will further lead to the highly stable Nb$_2O_5$ oxide formation as the formation enthalpy for Nb$_2O_5$ is -9.84 eV per Nb atom \cite{Massih2006}. However, the scenario changes drastically once Nb reaches its most preferred valence state 5+. The difference in energy between slabs containing Nb in subsurface and occupying O-top site is nearly 10 eV. This huge difference in energy will prevent Nb$^{5+}$ from escaping the oxide subsurface layer and forming Nb$_2O_5$. Stability of Nb$^{5+}$ in subsurface BaO slab and the huge energy difference between Nb$^{5+}$ at subsurface and surface O-top sites provide an unique opportunity to use Nb doped BaO as a support for Au single atoms where transfer of excess charge from Nb to Au can cause a strong binding of Au on the support.

\subsection{Adsorption of Au on Nb-Doped BaO Surface}

\subsubsection{Preferred binding site of single Au adatom}
\begin{figure}[h!]
\centering
\includegraphics[height=1.15in]{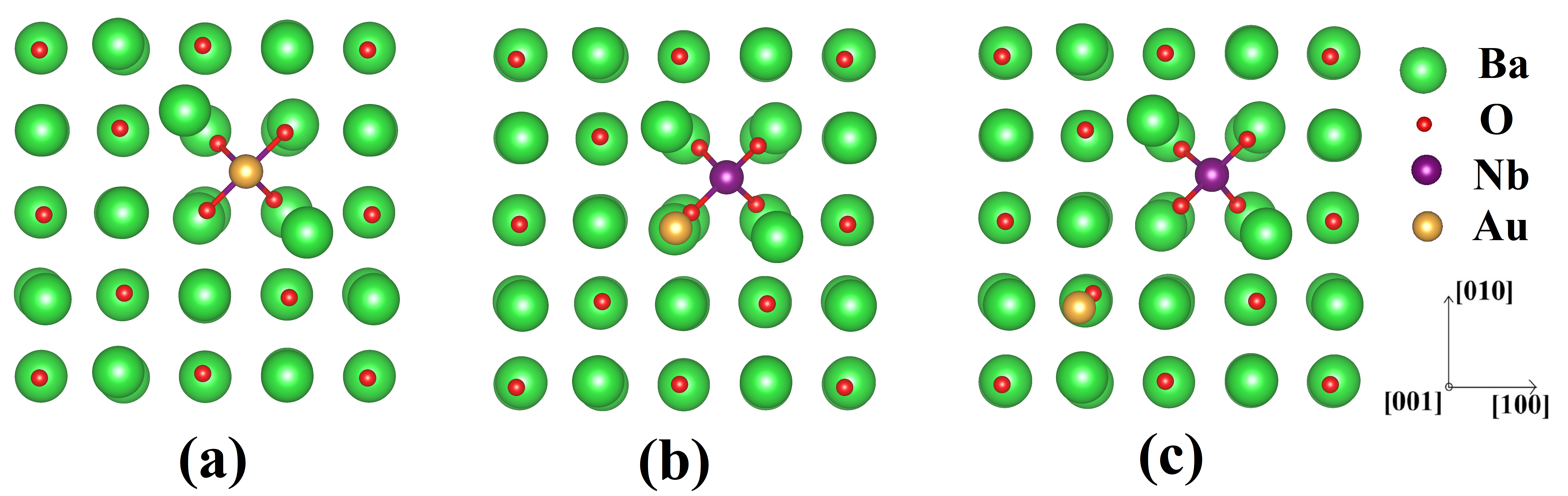}
\caption{Possible binding sites of a single Au atom on Nb-doped BaO\{001\} surface. Au occupies the (a) Nb top (b) near-Nb O-top and (c) far O-top sites respectively.} 
{\label{1_au}}
\end{figure}

The relaxed supercell that we used to calculate the energy of Nb doped at subsurface site, has been taken to understand the adsorption of Au. While calculating Au adsorption on one side of the doped BaO surface, necessary dipole corrections were included. Keeping in mind Hollow and O-top sites as the possible binding sites for noble metal atoms on pure BaO slab, we explored various possible binding sites for a Au atom on Nb-doped BaO slab. The potential binding sites are named as Nb-top, near-Nb O-top (when the binding oxygen atom is a part of the oxygen tetrahedra formed around Nb), and far O-top (when the binding oxygen atom is not a part of the tetrahedra formed around Nb) sites as shown in Fig.\ref{1_au}. Relative energies of Au single atom on these sites are listed in Table\ref{tab:binding}. We also explored hollow, Ba-top and bridge sites, however, for hollow site, the Au atom ends up at near-Nb O-top site, and in both the later cases the Au atom ended up on the neighbouring O-top sites indicating that these sites are not the preferred binding sites for Au atoms on Nb-doped BaO support. Our calculations revealed that Au atom binds most strongly on near-Nb O-top sites. For Au to get adsorbed on far O-top sites, it costs almost 0.29 eV more energy compared to near-Nb O-top sites. Surface Nb site is also higher in energy compared to the near Nb O-top sites. Hence oxygen atom near Nb, serves as the most preferred and strongest binding site for Au adsorption on Nb-doped BaO.

\begin{table}[h!]
\setlength{\tabcolsep}{4pt}
  \begin{center}
    \caption{Relative energies of Au single atom adsorbed at different sites on Nb-doped BaO\{001\} surface.}
    \label{tab:binding}
     \begin{tabular}{c|c}
      \multicolumn{1}{c|}{Binding Site} & \multicolumn{1}{c}{Relative energy (eV)}\\
      \hline
      Near-Nb O-top (Fig.\ref{1_au}b) & 0.00 \\
      Nb top (Fig.\ref{1_au}a) & 0.09\\
      Far O-top (Fig.\ref{1_au}c) & 0.29\\
   \end{tabular}
  \end{center}
\end{table}

\subsubsection{Stability of Au atoms}

\begin{table*}[t]
\setlength{\tabcolsep}{2pt}
  \begin{center}
    \caption{Adsorption energy ($|E_{ads}|$), Bader charge \textit{q} and average Au-Au bond length (d$_{Au-Au}$) for Au atom on Nb-doped and pure BaO \{001\} surfaces }
    \label{tab:binding}
     \begin{tabular}{c|c|c|c|c|c|c}
      \multicolumn{1}{c|}{} & \multicolumn{3}{c|}{Nb-doped BaO\{001\}} & \multicolumn{3}{c}{pure BaO\{001\}}\\
      \multicolumn{1}{c|}{No. of Au atoms} & \multicolumn{1}{c|}{$|E_{ads}|$ (eV)} & \multicolumn{1}{c|}{{\textit{q}}(e)} & \multicolumn{1}{c|}{$d_{Au-Au}$(\AA)} &	     \multicolumn{1}{c|}{$|E_{ads}|$ (eV)} & \multicolumn{1}{c|}{{\textit{q}}(e)} & \multicolumn{1}{c}{d$_{Au-Au}$(\AA)}\\
      \hline
      1 & 3.83 & -0.84 & & 1.88 & -0.40 &\\
      \hline
      2 & 3.78 & -0.83 & 4.64& 2.47 & -0.21 & 2.54\\
      \hline
      5 & 3.52 & -0.73 & 4.49 & 1.80& -0.23 & 2.74\\
   \end{tabular}
  \end{center}
\end{table*}

Stability of Au atom on pure and Nb-doped BaO supports were examined in terms of Au adsorption energy (per atom) for 1, 2 and 5 Au atoms using the equation 
\begin{equation}
E_{ads}=E_{sup+Au}-E_{sup}-nE_{Au}
\end{equation}

Here E$_{sup+Au}$, E$_{sup}$ and E$_{Au}$ refer to the energies of the support (pure or Nb-doped BaO) with Au atom, only support without Au and energy of Au single atom respectively and \textit{n} is the number of Au atoms adsorbed on the support. For adsorption of five Au atoms on Nb-doped BaO (001) slab, the adsorption sites are chosen in such a way that two Au atoms occupy the near-Nb O-top sites (strongest binding site) and the rest three Au atoms occupy the second nearest O-top sites. For pure BaO slab, adsorption energy of Au single atom is 1.88 eV and for 5 Au atoms it is 1.80 eV per Au atom. However, from Table\ref{tab:binding} it is evident that, compared to bare BaO surface, Nb-doped surfaces bind the Au single atom substantially well. The adsorption energy is almost 2 eV stronger when BaO is doped with Nb.

This strong adsorption of Au atom could be due to a charge transfer to Au. We have calculated the excess charge (difference in charge on the Au atom and number of valence electrons) on Au atom using Bader decomposition scheme,\cite{bader1990atoms, bader2007, bader2009} and the results are listed in Table\ref{tab:binding}. Our calculations revealed that there is a transfer of almost one electron from Nb to Au causing the Au atom to bind strongly on the support. Transfer of charge to Au is significantly low in case of pure BaO support. For Nb-doped BaO slab, Au atom acts as an acceptor of charge accepting excess electrons from Nb and the charge transfer process continues when number of adsorbed Au atoms is increased until a maximum of five Au atoms are adsorbed on the slab.

\subsubsection{Cluster formation of Au atoms}
In order to understand stability of adsorbed Au against segregation, we study interaction of two and five Au atoms adsorbed on pure BaO and  Nb-doped BaO surface. We find that the charged Au atoms resist clustering, one of the bottlenecks in fabricating single atom catalysts.
\begin{figure}[h!]
\centering
\includegraphics[height=1.4in]{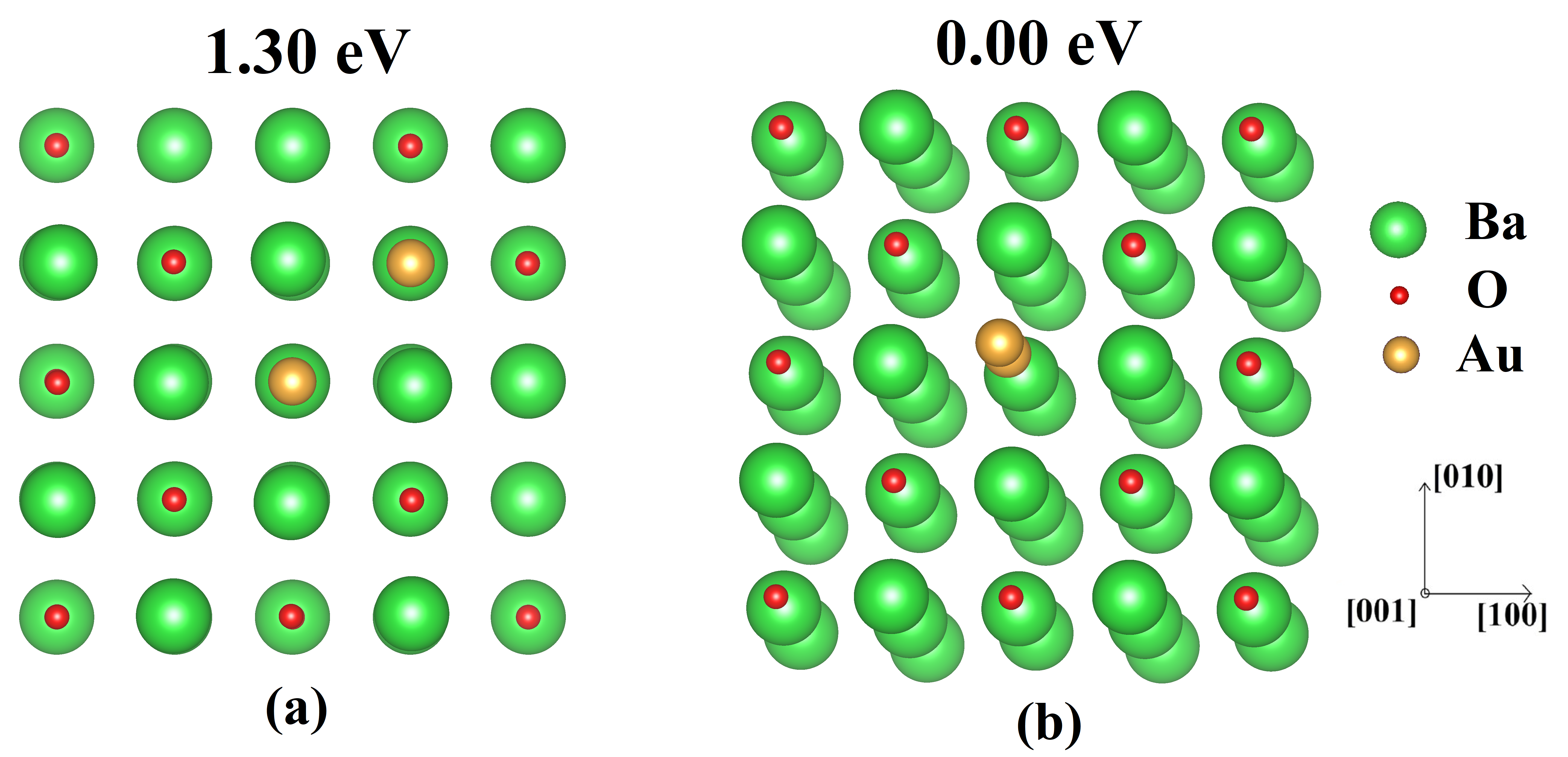}
\caption{Two Au atoms on pure BaO\{001\} surface and their relative energy.} 
{\label{bao2au}}
\end{figure}
\begin{figure}[h!]
\centering
\includegraphics[height=1.2in]{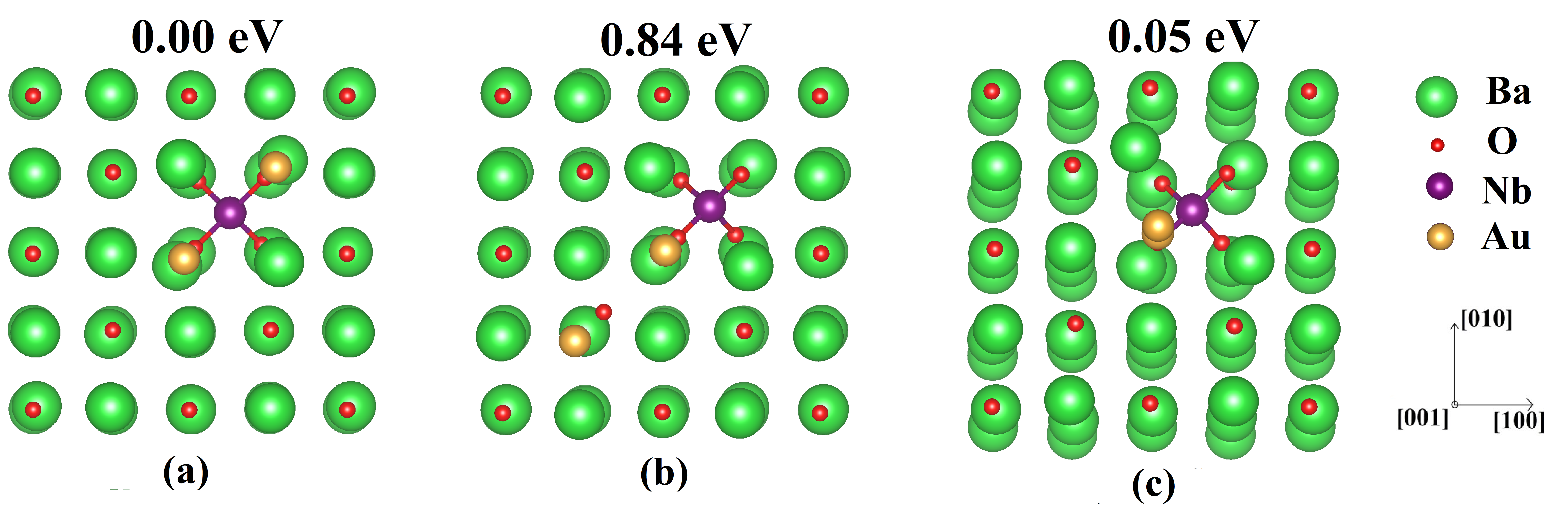}
\caption{Various possible arrangements of two Au atoms on Nb-doped BaO\{001\} support with their relative energies.} 
{\label{2au}}
\end{figure}
\begin{figure}[h!]
\centering
\includegraphics[height=1.1in]{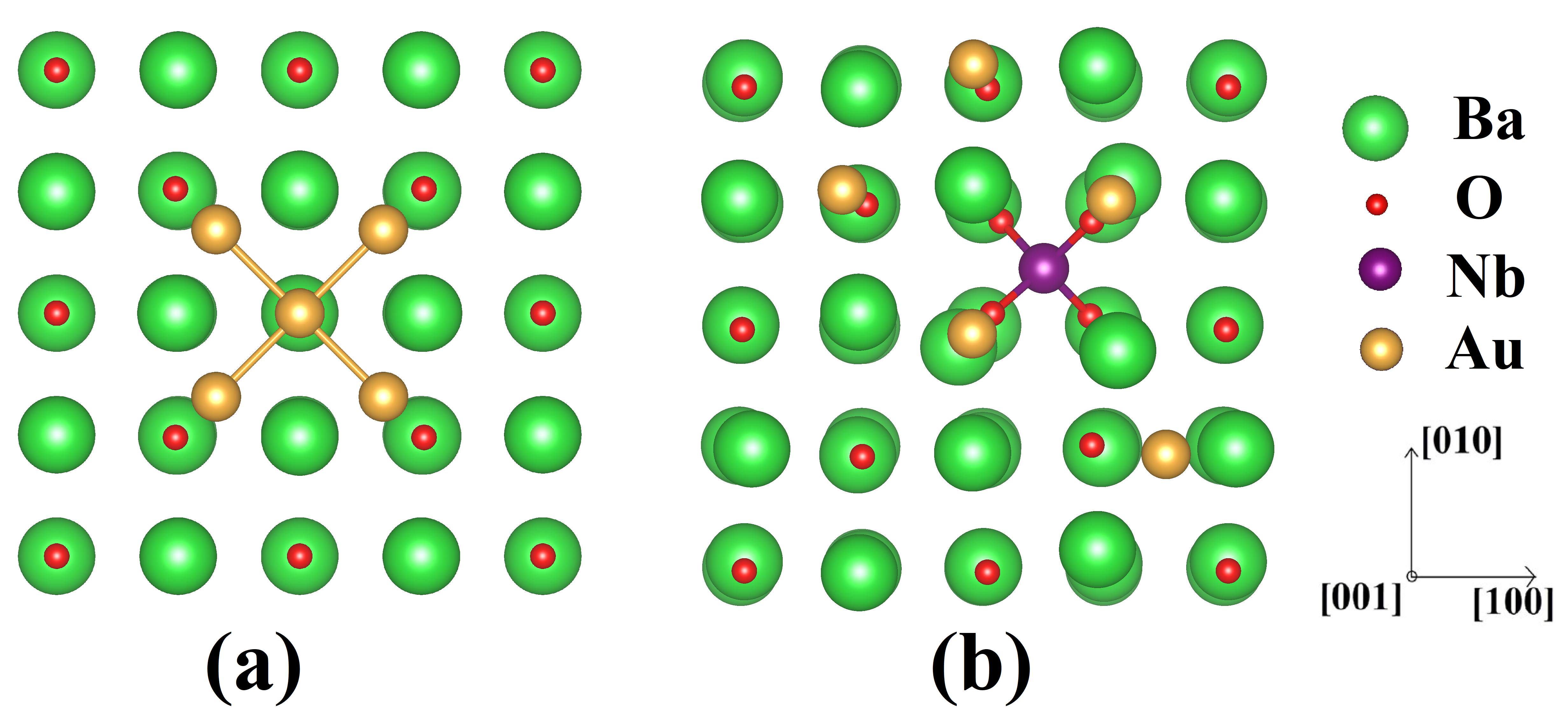}
\caption{Arrangement of five Au atoms on (a) pure and (b) Nb-doped BaO\{001\} surfaces} 
{\label{5au}}
\end{figure}
Fig.\ref{bao2au} shows the optimized structures for two Au atoms (a) on two different O-top sites and (b) together on same O-top site forming Au$_2$ cluster on pure BaO\{001\} surface, along with their relative energies. Comparison between these two arrangements revealed that formation of Au$_2$ cluster is more energetically favourable on pure oxide slab with a d$_{Au-Au}$ separation of 2.54 \AA, smaller than Au-Au bond length of 3.15 \AA, observed in Au clusters \cite{Zenge2015}. Similar trend has been observed for five Au SAs on pure BaO slab; Au atoms show the tendency to form cluster with an average Au-Au bondlength of 2.74\AA, as shown in Fig.\ref{5au}(a). Hence, Au atoms bound on pure BaO slab are prone to form larger particles.

On the other hand, for Nb-doped BaO slab, a completely opposite trend has been observed. The Au atoms anchored on the doped oxide surface preferred to remain atomically dispersed and resist cluster formation. Figure \ref{2au} depicts the possible configurations taken by two Au atoms on Nb-doped BaO slab: the optimized structures for two Au atoms (a) on two near-Nb O-top sites (b) one Au on a near-Nb O-top sites and the other one on far O-top site. Figure\ref{2au}(c) is the initial structure of two Au atoms adsorbed on the same O-top sites. The relative energies between these configurations are also mentioned in Fig.\ref{2au}.  
We find that, Fig.\ref{2au}(a) is the most favoured configuration for two Au atoms on Nb-doped BaO support with a bond length of 4.64\AA, which is almost 1.5\AA larger than the maximum bond length reported (3.15\AA) \cite{Zenge2015}. The third configuration (\ref{2au}c) is not stable and after optimization it comes to configuration (\ref{2au}a). This clearly shows that the Nb-doped BaO \{001\} surface serves as an agglomeration-resistant support for Au SAs. For five Au SAs anchored on Nb-doped BaO (Fig.\ref{5au}(a)), the Au atoms seem to move away from each other indicating that they resist forming larger particle and prefer to stay dispersed in the form of single atoms. Hence, it is evident that Nb-doping not only binds Au SAs more strongly but also prevents cluster formation of the SAs.

\subsubsection{Electronic structure}
\begin{figure*}
\centering
\includegraphics[height=3.5in]{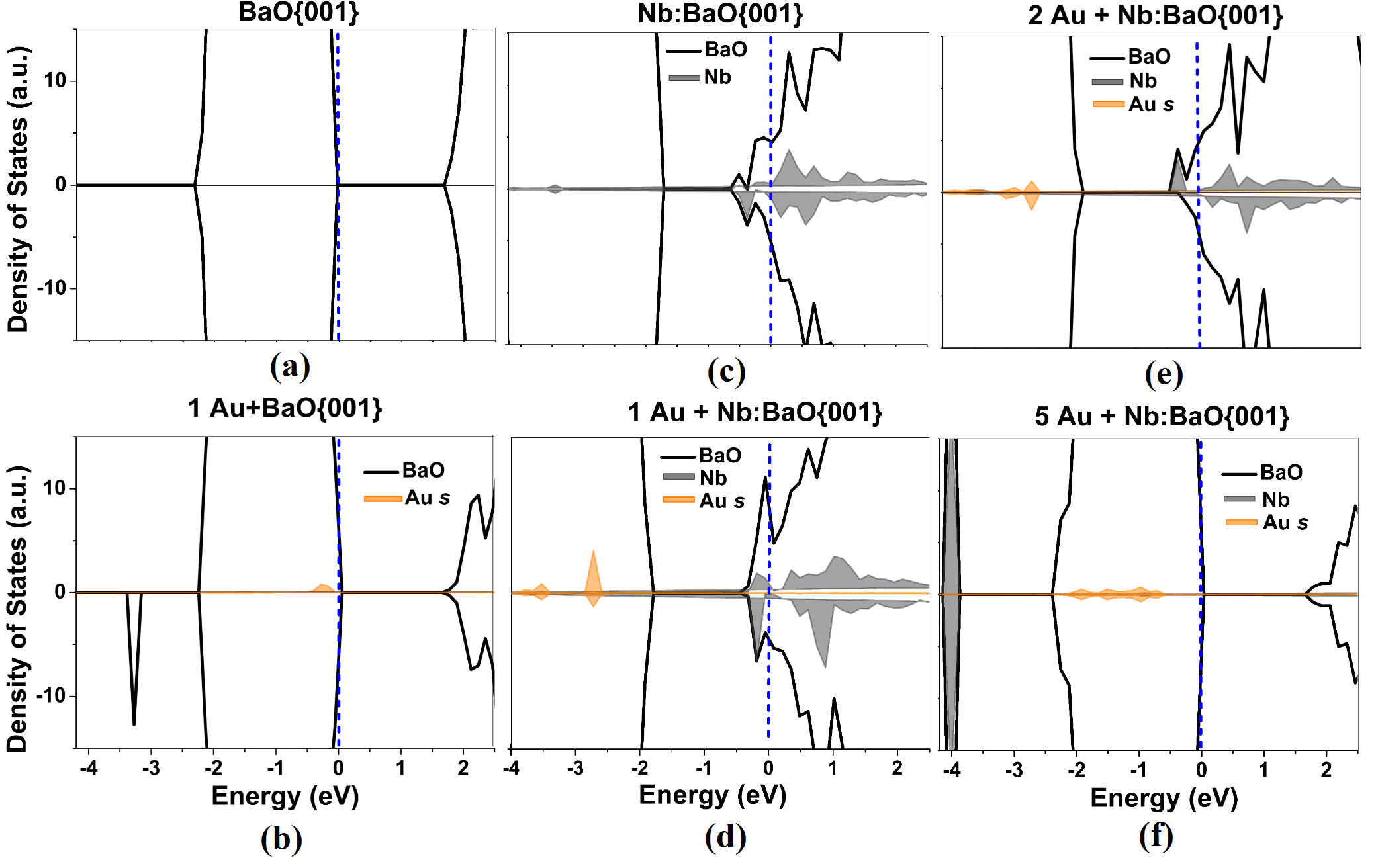}
\caption{Atom-resolved density of states for (a) pure BaO (b) 1 Au atom adsorbed on pure BaO and (c) Nb-doped BaO\{001\} surface; (d) 1 Au, (e) 2 Au and (f) 5 Au atoms adsorbed on Nb-doped BaO\{001\} surface. For Au, only 6\textit{s} states are shown. The blue dashed line indicates the Fermi level.} 
{\label{total-dos}}
\end{figure*}

To explain why charge transfer takes place between Nb and Au atoms, we calculate density of states (DOS) for (a) pure BaO (b) 1 Au atom adsorbed on pure BaO, (c) Nb-doped BaO, and (d) 1 Au, (e) 2 Au, and (f) 5 Au atoms adsorbed on Nb-doped BaO\{001\} surfaces. Atom resolved DOS are shown in Fig.\ref{total-dos}; for Au atoms only 6\textit{s} states are shown. Being an insulator, pure BaO\{001\} surface shows no defect states in the band gap region, and a single atom Au adsrobed on pure BaO has its 6\textit{s} orbital half-filled. Once Nb is doped in BaO, midgap states, mostly carrying the d-orbital signature of Nb appears in the bandgap region of BaO. These mid gap states eases the charge transfer between Nb and Au. For BaO doped with Nb, the highest occupied level of the system is pushed to CBM of BaO. Once one or two single atom of Au gets adsorbed on Nb doped BaO support, the highest occupied level of the system remains in CBM with both the 6\textit{s} states of Au now being occupied and degenerate, which is attributed to the charge transfer from Nb to Au. It is only after the adsorption of 5 Au atoms, the highest occupied level comes down to VBM, indicating that all the excess 5 electrons are transferred to Au from Nb. Unlike pure BaO support, the charge transfer between Nb and Au is evident for the doped one. A single Au can accommodate one extra electron from Nb in its \textit{s} orbital and this charge transfer process continues till all 5 Au atoms are adsorbed on the support accepting one electron each from Nb. This in turn helps Nb to attain the most preferred charge state, 5+. A careful observation of the Fermi level position (blue dashed line) in DOS plots reveals that the Fermi level of the system which was initially near CBM, comes down to VBM of BaO when all the excess five electrons from Nb are donated to 5 Au atoms, indicating that 5 Au atoms bound on BaO\{001\} slab with subsurface Nb interstitial is a well stable system as a whole.

\section{Discussion}
Charge transfer from substitutional dopant to adsorbate for stabilizing single atom of Au has been reported earlier \cite{prada2013}. However, two factors limit their effectiveness in charge transfer, as noted by Stavale et. al. \cite{Stavale2012}: 1)  change in the charge state of substitutionally doped transition metal in the host, 2) the electron that is to be transferred to the adsorbate can also get captured by cationic vacancies or morphological defects in the doped oxide. Our approach mitigates these limitations. Nb doped at substitutional site in BaO would be able to transfer 3 electrons, while at interstitial site it will be able to transfer 5 electrons. As implantation can be done on single crystalline BaO, creation of cationic vacancies or morphological defects could be minimized. 

To our surprise, stability against segregation of Au when it accepts charge from the doped TM have not been shown in any of the previous studies. Our results explain why formation of well dispersed two dimensional Au layer on Mo doped CaO is favourable, as charge gets transferred from Mo to Au, while in case of Cr doped MgO, a 3D island of Au forms, as Cr could not transfer any charge to Au \cite{Stavale2012}. 

Based on our calculations we propose following experimental steps to achieve well bound single atom Au on Nb-doped BaO. First, single layer of Au can be dispersed on \{001\} surface of single crystal BaO followed by implantation of Nb ions into the BaO surface. Once Nb gets doped into BaO at the subsurface layer, it transfers charge to Au and binds few of those gold SAs on near Nb surface oxygen sites of BaO. The loosely bound Au could be washed away leaving only strongly bound single Au atoms.

\section{Conclusions}
Based on first principles calculations, we propose a new way to form single atom Au, supported on BaO. We show that such charged Au atoms are also stable against agglomeration, which explains already reported formation of 2D gold layer on Mo doped CaO. To achieve such well dispersed Au, Nb needs to be implanted in BaO which is stable at interstitial site in BaO. This process of stabilizing Au on an oxide supports provides a distinct advantage over the known methods involving substitutional doping of oxide with transition metals using conventional chemical routes. Transition metal at interstitial site transfers more charge to adsorbate on oxide surface. Also ion-implantation provides area- and concentration-controlled doping as it is a well established doping route in electronic industries to create \textit{n} and \textit{p} type semiconductors. Although our results deal with Nb doped BaO for binding Au single atoms, it opens up the possibility to explore other combination of oxides, doped with different transitions metals to bind noble metal single atoms. 

\begin{acknowledgements}
 D.M. gratefully acknowledges the support from Indo-German Centre for Sustainability (IGCS) research for providing the opportunity to visit Technical University of Munich for three months and carry out most of this work, and also the institute post-doctoral fellowship provided by IIT Madras, India. D.M. and S.K.Y. acknowledge Dr. Somnath Bhattacharya's help for providing access to VASP source code.
\end{acknowledgements}

\bibliographystyle{ieeetr}
\bibliography{paper}

\end{document}